\documentclass[prx,nofootinbib,twocolumn,superscriptaddress]{revtex4-2}
\usepackage[colorlinks=true, linkcolor=blue, urlcolor=blue]{hyperref}

\usepackage{lipsum}

\usepackage{amsmath,amssymb,bm,graphicx}
\usepackage{multirow}
\usepackage{dcolumn}
\usepackage{latexsym}
\usepackage[normalem]{ulem}
\usepackage{orcidlink}

\usepackage{comment}
\def\<{\langle}
\def\>{\rangle}

\newcommand{\tr}{\mathrm{Tr}}

\newcommand{\Exp}[1]{\langle #1 \rangle}

\newcommand{\HA}{\mathop{\mathcal{H}}\nolimits}

\newcommand{\I}{\mathop{\mathbb{I}}\nolimits}
\newcommand{\R}{\mathop{\mathbb{R}}\nolimits}
\newcommand{\CA}{\mathop{\mathbb{C}}\nolimits}
\newcommand{\ketbra}[2]{| #1 \rangle \langle #2 | }
\newcommand{\bracket}[1]{\langle #1  \rangle}

\newcommand{\ket}[1]{| #1 \rangle}

\newtheorem{Theorem}{Theorem}

\newtheorem{Corollary}{Corollary}

\newtheorem{Conj}{Conjecture}

\newcommand{\lM}{\lambda_{\rm M}}
\newcommand{\lm}{\lambda_{\rm m}}
\newcommand{\lSm}{\lambda_{\rm sm}}

\newcommand{\dc}{\textcolor{blue}}

\begin{document}

\title{Uncertainty relations based on state-dependent norm of commutator}

\author{Aina Mayumi}
\thanks{a.mayumi1441@gmail.com}
\affiliation{College of Systems Engineering and Science, Shibaura Institute of Technology, Saitama 330-8570, Japan}

\author{Gen Kimura\,\orcidlink{0000-0003-4288-2024}}
\thanks{gen@shibaura-it.ac.jp [Corresponding author]}
\affiliation{College of Systems Engineering and Science, Shibaura Institute of Technology, Saitama 330-8570, Japan}

\author{Hiromichi Ohno \,\orcidlink{0000-0001-5498-3311}}
\thanks{h\_ohno@shinshu-u.ac.jp}
\affiliation{Department of Mathematics, Faculty of Engineering, Shinshu University, 4-17-1 Wakasato, Nagano 380-8553, Japan}

\author{Dariusz Chru\'sci\'nski\,\orcidlink{0000-0002-6582-6730}}
\thanks{darch@fizyka.umk.pl}
\affiliation{Institute of Physics, Faculty of Physics, Astronomy and Informatics  Nicolaus Copernicus University, Grudzi\c{a}dzka 5/7, 87--100 Toru\'n, Poland}

\pacs{}

\begin{abstract}
We introduce two uncertainty relations based on the state-dependent norm of commutators, utilizing generalizations of the B\"ottcher-Wenzel inequality. The first relation is mathematically proven, while the second, tighter relation is strongly supported by numerical evidence. Both relations surpass the conventional Robertson and Schr\"odinger bounds, particularly as the quantum state becomes increasingly mixed. This reveals a previously undetected complementarity of quantum uncertainty, stemming from the non-commutativity of observables. We also compare our results with the Luo-Park uncertainty relation, demonstrating that our bounds can outperform especially for mutually unbiased observables.
\end{abstract}

\maketitle


\section{Introduction}


The uncertainty principle is a fundamental characteristic of quantum mechanics and has a rich history \cite{Jammer1989Quantum}.
Beginning with Heisenberg's initial exploration using the gamma-ray microscope thought experiment \cite{Heisenberg1927}, Kennard \cite{Kennard1927Quantenmechanik}, Wyle \cite{Weyl1928} and Robertson \cite{Robertson1929Uncertainty} established the relation expressing the uncertainty by the standard deviation.
Specifically, Robertson showed, for any observables $A$ and $B$ represented by Hermitian operators, the uncertainty relation
\begin{align}\label{RU}
	V_\rho(A) V_\rho(B) \ge \frac{1}{4}|\langle [A,B]\rangle_\rho |^2,
\end{align}
where $\langle X \rangle_\rho := \tr X \rho$ and $V_\rho(X):= \tr (X-\langle X \rangle_\rho)^2 \rho$ are the expectation value and the variance (squared standard deviation) for an observable $X$ under a quantum state $\rho$, and $[A,B]:= AB - BA$ denotes the commutator of $A$ and $B$.
The Robertson relation \eqref{RU} elegantly illustrates a fundamental trade-off between the uncertainties of non-commutative observables, highlighting the inherent connection between non-commutativity and uncertainty in quantum mechanics.
Shortly after this formulation, Schr\"odinger derived a tighter inequality \cite{schrodinger1930}
\begin{align}\label{SU}
	V_\rho (A) V_\rho (B) \ge \frac{1}{4}\Bigl|\langle [A,B] \rangle_\rho \Bigr|^2 + C_\rho(A,B),
\end{align}
with
\begin{align}\label{SUC}
	C_\rho(A,B):=  \Bigl|\frac{1}{2}\langle \{A,B\}\rangle_\rho -\langle {A}\rangle_\rho \langle{B}\rangle_\rho \Bigl|^2,
\end{align}
where $\{A,B\}:= AB + BA$ denotes the anti-commutator of $A$ and $B$.
The additional term \eqref{SUC} corresponds to the (symmetrized) covariance between $A$ and $B$.
Hence, the Schr\"odinger relation \eqref{SU} improves upon the Robertson relation \eqref{RU} by the amount of ``classical" covariance.

Since the original formulation by these pioneers, uncertainty relations have been extensively studied by many researchers, both for preparation and measurement uncertainties (see, e.g., \cite{busch2016quantum}).
Given the vast amount of literature on this topic, we briefly highlight only a few results focusing on preparation uncertainty relations.
To name a few, some authors are investigating on the sums of variances \cite{Maccone,Wang,SONG20162925,Fan2020}, or the uncertainty regions \cite{Li2015,Busch2019,Zhang2021}.
Other measures of uncertainty have also been employed, such as entropies \cite{Deutsch,Maassen,Kraus_1987,Berta_2010,RevModPhys.89.015002}, Wigner-Yanase skew information \cite{LZ2004,Luo2,Yanagi_2005,LZ2005,Kosaki_2005,Li_2009,FURUICHI2009179,Yanagi_2010,Chen_2016}, the maximum probabilities \cite{Landau1961,MiyaderaImai}, Fisher information \cite{Gibilisco_2007,FSG2015,CG2022,TGFF2022}, and quantum coherence \cite{SPB2016,Liu2016,Plenio2,PhysRevA.96.032313,Rastegin2017,Luo_2019}. For reviews of uncertainty relations, we refer \cite{Peres1995,Wehner_2010,busch2016quantum,HilgevoordUffink2024,Englert}.

Focusing on the relations for the product of variances, Luo \cite{Luo2} and Park \cite{Park2005} independently obtained an interesting relation which generalizes the Robertson relation:
\begin{align}\label{eq:Luo}
	V_\rho(A) V_\rho(B) \ge \frac{1}{4}|\langle [A,B] \rangle_\rho|^2 + C_\rho(A)C_\rho(B).
\end{align}
Here, $C_\rho(X)$ represents a ``classical" uncertainty of an observable $X$ under a state $\rho$ defined by
$$
C_\rho(X):= V_\rho(X) - I_\rho(X), 
$$
where
$$
I_\rho(X):= -\frac{1}{2}\tr [\sqrt{\rho},X]^2 = \tr X^2 \rho - \tr \sqrt{\rho} X \sqrt{\rho} X
$$
is the Wigner-Yanase skew information \cite{wigner1963}. In the following, we refer to \eqref{eq:Luo} as the Luo-Park (LP) relation.
One observes an apparent resemblance between the Schr\"odinger relation \eqref{SU} and the LP relation \eqref{eq:Luo}, but they exhibit fairly different behavior, as we shall see below.

In this paper we derive another uncertainty relations for the product of variances by generalizing the B\"ottcher-Wenzel (BW) inequality \cite{Bottcher2005} (cf. also \cite{VongJin2008BottcherWenzel,BOTTCHER20081864,Audenaert2010VarianceBounds})
\begin{align}\label{BW}
	\|[A,B]\|^2 \le 2 \|A\|^2 \|B\|^2 ,
\end{align}
where $\|A\|:= \sqrt{\tr A^\dagger A}$ stands for the Frobenius norm for a (Hilbert-Schmidt class) linear operator \cite{BhatiaMatrixAnalysis}. (Throughout the paper, the symbol $\|\cdot \|$ represents the Frobenius norm for linear operators.)
The BW inequality \eqref{BW} is readily proven when either $A$ or $B$ is normal, yet it becomes quite non-trivial if neither of them is normal. (For reader's convenience, we include the proof of the BW inequality for the cases of normal operators in Appendix \ref{app:BW}).
We refer to \cite{Audenaert2010VarianceBounds} for an elegant proof of the inequality, which employs a quantum information technique.
The BW inequality is tight, indicating that non-zero matrices $A$ and $B$ always exist for which equality is attained\footnote{Interestingly, if at least one of operator $A$ or $B$ is positive, then $\|[A,B]\|^2 \le  \|A\|^2 \|B\|^2$, and if both are positive, then $\|[A,B]\|^2 \le 1/2 \|A\|^2 \|B\|^2$ \cite{Bottcher2005,Bhatia2008,BOTTCHER20081864}. }. In other words, the factor $2$ appearing in \eqref{BW} cannot be further improved.
Notice that this holds true even when $A$ and $B$ are restricted to being Hermitian.

The BW inequality, initially sparked by pure mathematical interest, has only recently begun to be recognized among the physics community. To the best of the authors' knowledge, it has been utilized only in \cite{Kimura2017Universal,Luo_2019,CKKS2021,SzczyAlicki,Caravelli2021,Koczor2021} for applications in quantum physics.
In particular, it has been directly used in \cite{Kimura2017Universal,CKKS2021} to derive a universal constraint between relaxation times \cite{PhysRevA.66.062113,CKKS2021,CHRUSCINSKI2021293} reflecting the completely positivity condition in quantum Markovian dynamics \cite{Gorini1976CPDS,cmp/1103899849}.
The BW inequality (with one operator being positive) was also used in \cite{Luo_2019} to derive an uncertainty relation for a quantum coherence.

Given that the BW inequality \eqref{BW} includes a commutator term, one would be interested in its application to uncertainty relations. Indeed, it immediately implies the following uncertainty relation for a maximally mixed state $\rho_{\max} := \frac{\I}{d}$
\begin{align}\label{BWU}
	V_{\rho_{\max}}(A) V_{\rho_{\max}}(B) \ge \frac{1}{2d^2} \|[A,B]\|^2.
\end{align}
This can be shown by applying $\hat{A}:=A- \Exp{A} \I$ and $\hat{B}:=B- \Exp{B} \I$ to the BW inequality by noting that $V_{\rho_{\max}}(A)  = \frac{\|\hat{A} \|^2}{d},V_{\rho_{\max}}(B) = \frac{\|\hat{B}\|^2}{d}$, and $[\hat{A},\hat{B}] = [A,B]$.
Interestingly, \eqref{BWU} is already stronger than \eqref{RU} and \eqref{SU} in particular situations.
For instance, consider the qubit case $(d=2)$ and take $A=\sigma_x$ and $B=\sigma_y$. Then both (\ref{RU}) and (\ref{SU}) give trivial bound $V_{\rho_{\max}}(\sigma_x) V_{\rho_{\max}}(\sigma_y)  \geq 0$, whereas \eqref{BWU} implies $V_{\rho_{\max}}(\sigma_x) V_{\rho_{\max}}(\sigma_y) \geq 1$. 

The main idea of the present paper is to generalize \eqref{BWU} beyond maximally mixed states.
Recently, we explored several generalizations of the BW inequality using a state-dependent norm \cite{MKHC}.
It turns out that the generalized BW inequalities allow us to derive the following uncertainty relations:
\begin{align}\label{newULoos}
	{V_\rho(A) V_\rho(B)} \ge \frac{\lm^2}{2\lM} \|[A,B]\|^2_\rho
\end{align}
and
\begin{align}\label{newU}
	{V_\rho(A) V_\rho(B)}\ge \frac{\lm \lSm}{\lm + \lSm} \|[A,B]\|^2_\rho ,
\end{align}
where $\|A\|_\rho:= \sqrt{\tr (A^\dagger A \rho)}$ defines the state-dependent Frobenius semi-norm, and $\lm, \lSm$, and $\lM$ denote the smallest, the second smallest, and the largest eigenvalues of $\rho$, respectively.
While the first relation \eqref{newULoos} is derived from a generalized BW inequality that has already been mathematically proven, the second relation \eqref{newU} is based on a generalization of the BW inequality that is conjectured to be the tightest and is strongly supported by numerical optimization \cite{MKHC}.
In this paper, we conduct a systematic and thorough comparison between our relations \eqref{newULoos}, \eqref{newU}, and the Robertson, Schr\"odinger, and also LP relations.
In particular, we analytically compute the averaged bounds for all these relations in qubit systems and show that our bounds surpass the Robertson and Schr\"odinger bounds as a state becomes more mixed.
Observing that the bounds in \eqref{newULoos} and \eqref{newU} are composed of the commutator between observables, thus, our relations are detecting a trade-off of non-commutative observables that has been unidentified by conventional uncertainty relations.
On the other hand, we observe that the LP bound outperforms our bounds on average. We next compare the cases of mutually unbiased observables, or traditionally a complementary pair of observables \cite{Schwinger,Ivonovic_1981,WOOTTERS1989363}. We will show that our relation \eqref{newU} can outperform the LP bound for the averaged bounds for all mutually unbiased observables. This fact might be interesting because uncertainty relations are prominently manifested in complementary physical quantities.



The paper is organized as follows: In Section \ref{S-II}, we provide a brief review of generalizations of the original BW inequality \eqref{BW}. 
As their application, the generalized uncertainty relations \eqref{newULoos} and \eqref{newU} are introduced in Section \ref{S-III}.
Section \ref{S-IV} offers a thorough comparison of these relations with the Robertson and Schr\"odinger relations, as well as the LP bound, in two cases: for qubit systems (Sec.\ref{AvOb}) and for mutually unbiased observables (Sec.\ref{AvMUOb}). Finally, we summarize our findings in Section \ref{S-IV}.

\section{Extending the B\"ottcher-Wenzel inequality} \label{S-II}

In this section, we briefly review the discussion in \cite{MKHC}, adjusting it with the application to the uncertainty relation in mind.
In this paper, we restrict ourselves to finite-level quantum systems associated with a $d$-dimensional complex Hilbert space $\HA \simeq \CA^d$. We use the standard Dirac notation with (ket) vectors $\ket{\psi},\ket{\phi} \in \HA$, where their inner product and the norm are denoted by $\bracket{\psi,\phi}$ and $\|\psi\| := \sqrt{\bracket{\psi,\psi}}$, respectively. Note that $\ketbra{\psi}{\phi}$ is a linear operator. The adjoint matrix of a linear operator $A$ is denoted by $A^\dagger$.
However, $n$-dimensional real vectors in $\R^n$ are distinctly denoted as ${\bm a} = (a_1,a_2,\ldots,a_n), {\bm b} = (b_1,b_2,\ldots,b_n)$, where the Euclidean inner product and norm are denoted by $|{\bm a}| := \sqrt{\sum_{i=1}^n a_i^2}$, ${\bm a}\cdot {\bm b} := \sum_{i=1}^n a_ib_i$, respectively.
We also use the notation $\overline{\alpha}$ and $|\alpha|$ to represent the complex conjugate and absolute value of a complex number $\alpha$, respectively.


For any (not necessarily normalized) density matrix $\rho$, we define a semi-inner product between matrices $A,B$ by
\begin{align}\label{WIP}
	\langle A,B \rangle_\rho := \tr (A^\dagger B \rho).
\end{align}
By the positive semi-definiteness of $\rho$ and the cyclic property of the trace operation, it is straightforward to see (i) $\langle A,A\rangle_\rho \ge 0$, (ii) $\overline{\langle B,A \rangle_\rho} = \langle A,B \rangle_\rho $, and (iii) $\langle A, \alpha B + \beta C \rangle_\rho = \alpha \langle A,  B\rangle_\rho + \beta \langle A, C \rangle_\rho $ for all matrices $A,B,C$ and $\alpha,\beta \in \CA$.
The weighted Frobenius semi-norm is given as the induced semi-norm:
\begin{align}\label{Wnorm}
	\| A \|_\rho := \sqrt{\langle A,A \rangle_\rho} = \sqrt{\tr (A^\dagger A \rho)},
\end{align}
which satisfies (i) $\|A\|_\rho \ge 0$, (ii) $\|\alpha A\|_\rho = |\alpha | \|A\|_\rho $, and (iii) $\| A + B\|_\rho \le \| A \|_\rho  + \|B\|_\rho $ for all matrices $A,B$ and $\alpha \in \CA$.
The expressions \eqref{WIP} and \eqref{Wnorm} generalize the Hilbert-Schmidt inner product and Frobenius norm, respectively, when $\rho = \I$.

Let $\lambda_i$ ($i = 1, \ldots, d$) the eigenvalues of $\rho$ arranged in descending order: $0 \le \lambda_1 \leq \lambda_2 \leq \cdots \leq \lambda_d$ with the corresponding normalized eigenvectors $\ket{\lambda_i}$. Throughout this paper, the notations
$$
\lm = \lambda_1,\lSm= \lambda_2 \ {\rm and} \ \lM = \lambda_d
$$
represent the smallest, the second smallest and the largest eigenvalues of $\rho$, respectively. Using the eigenvalue decomposition $\rho = \sum_i \lambda_i \ketbra{\lambda_i}{\lambda_i}$, one has $\|A\|^2_\rho = \tr A^\dagger A \rho = \sum_i \lambda_i \bracket{\lambda_i|A^\dagger A \lambda_i}$. Since $\|A\|^2 = \tr A^\dagger A = \sum_i \bracket{\lambda_i|A^\dagger A \lambda_i}$, we observe
\begin{align}
	\lm \|A\|^2 \le \|A\|^2_\rho \le \lM \|A\|^2.
\end{align}
Therefore, we get
\begin{align}\label{LoB}
	\|[A,B]\|^2_\rho &\le \lM \|[A,B]\|^2 \nonumber \\
	&\le 2 \lM \|A\|^2 \|B\|^2 \le \frac{2 \lM}{\lm^2} \|A\|^2_\rho \|B\|^2_\rho,
\end{align}
where in the second inequality we have used the BW inequality \eqref{BW}.
According to the numerical simulations, however, the bound \eqref{LoB} appears not to be tight and could be further improved.
In \cite{MKHC}, we proposed a conjecture for the tight bound:
\begin{Conj}\label{conj}
	For any positive definite matrix $\rho$ and for any complex matrices $A,B$, we have
	\begin{equation}\label{conj:BW}
		\|[A,B]\|^2_\rho \le \frac{\lm + \lSm}{\lm \lSm} \|A\|^2_\rho \|B\|^2_\rho.
	\end{equation}
	This inequality is sharp, meaning that there exist non-zero matrices $A,B$, in particular Hermitian matrices, that achieve equality.
\end{Conj}
Several remarks are in order:
First, unlike the BW inequality, this bound is far from trivial even for Hermitian matrices. This is mainly because there are essentially three non-commutative matrices involved: $A$, $B$, and $\rho$.
Second, we have conducted the following numerical optimization
$$
\max_{A,B \neq 0} \frac{\|[A,B]\|^2_\rho}{\|A\|^2_\rho \|B\|^2_\rho}
$$
for randomly generated positive matrix $\rho$ up to size $d=15$ and the results perfectly match the conjectured bound $\frac{\lm + \lSm}{\lm \lSm}$ in \eqref{conj:BW}.
Third, the bound has been proven for $d=2$, hence can be applied to qubit systems.
In Appendix \ref{app:Pd2}, we provide an independent proof from the one given in \cite{MKHC}, adjusted for Hermitian matrices.
Fourth, for any $\rho$, there are non-zero Hermitian matrices $A$ and $B$, e.g., $A=\lambda_2 |\lambda_1 \rangle \langle \lambda_1|-\lambda_1|\lambda_2 \rangle\langle \lambda_2|$ and $B=|\lambda_1 \rangle \langle \lambda_2|+|\lambda_2 \rangle \langle \lambda_1|$, that achieve equality in \eqref{conj:BW}, thus proving the tightness part of \eqref{conj:BW}.
Fifth, the bounds \eqref{conj:BW} (and also \eqref{LoB}) are generalizations of the BW inequality \eqref{BW} when $\rho = \I$ noting that $\lm= \lSm (= \lM) = 1$.
Finally, the bound can be trivially extended to include positive-semidefinite matrices $\rho$ by interpreting the bound in \eqref{conj:BW} as infinite when $\lm$ equals zero.

\section{Application to uncertainty relations}   \label{S-III}

The proposed uncertainty relations \eqref{newULoos} and \eqref{newU} can be readily shown by the bounds \eqref{LoB} and \eqref{conj:BW} as follows.
For any pair of observables $A$ and $B$, their variances are expressed by using semi-norms:
$$
V_\rho(A) = \|\hat A\|^2_\rho, \ \ \ \ V_\rho(B)= \|\hat{B}\|^2_\rho,
$$
where $\hat A := A - \langle A \rangle_\rho \I, \hat B := B - \langle B \rangle_\rho\I$.
Therefore, we derive relations \eqref{newULoos} and \eqref{newU} by applying $\hat{A}$ and $\hat{B}$ to \eqref{LoB} and \eqref{conj:BW} with a quantum state $\rho$ and noting $[\hat A,\hat B] = [A,B]$.
Summarizing, we have
\begin{Theorem}
	For a $d$-level quantum system, the following uncertainty relation holds between observables $A$,$B$ under a state $\rho$:
	\begin{align}\label{newULoos2}
		V_\rho(A) V_\rho(B) \ge \frac{\lm^2}{2\lM} \|[A,B]\|^2_\rho.
	\end{align}
	Here $\lm$ and $\lM$ are the smallest and the largest eigenvalues of $\rho$.
\end{Theorem}
Moreover, based on Conjecture \ref{conj}, which is substantiated by numerical computations and proved for qubit system, we have
\begin{Conj}
	For a $d$-level quantum system, the following uncertainty relation holds between observables $A$,$B$ under a state $\rho$:
	\begin{align}\label{newU2}
		V_\rho(A) V_\rho(B) \ge \frac{\lm \lSm}{\lm + \lSm} \|[A,B]\|^2_\rho.
	\end{align}
	Here $\lm$ and $\lSm$ are the smallest and the second smallest eigenvalues of $\rho$.
\end{Conj}
For qubit systems one has therefore
\begin{Corollary} For a $2$-level quantum system, the following uncertainty relation holds
	\begin{align}\label{}
		V_\rho(A) V_\rho(B) \ge \lambda_1 \lambda_2 \|[A,B]\|^2_\rho ,
	\end{align}
	where $\lambda_1 \leq \lambda_2$ are eigenvalues of $\rho$.
\end{Corollary}
For a pure qubit state $\lambda_1=0$ and hence the above bound is trivial. However for a genuine mixed state the bound is nontrivial for any pair of non-commuting observables. 

The bounds in both relations are composed of the commutator, so our relations resemble the Robertson relation. However, as we will see below, they are stronger than the Robertson relation (and even than the Schr\"odinger relation) as a state becomes more mixed.

In the following, we present a detailed comparison of the uncertainty relations \eqref{newULoos2} and \eqref{newU2} with the Robertson and Schr\"odinger relations \eqref{RU} and \eqref{SU}, as well as the LP relation \eqref{eq:Luo}, specifically within the context of a qubit system.
The lower bounds of each uncertainty relation, in the order of Robertson, Schr\"odinger, Luo-Park, \eqref{newULoos2} and \eqref{newU2} are given respectively by
\begin{subequations}\label{AllBdd}
	\begin{align}
		&B_{\rm R}(A,B,\rho) = \frac{1}{4}\Bigl|\Exp{[A,B]}_\rho\Bigr|^2, \\
		&B_{\rm S}(A,B,\rho) = \frac{1}{4}\Bigl|\Exp{[A,B]}_\rho\Bigr|^2+ \Bigl|\frac{1}{2}\Exp{\{A,B\}}_\rho-\Exp{A}_\rho\Exp{B}_\rho\Bigl|^2, \\
		&B_{\rm LP}(A,B,\rho) = \frac{1}{4}\Bigl|\Exp{[A,B]}_\rho\Bigr|^2 + \nonumber \\
		&\quad \Bigl( \tr \sqrt{\rho} A \sqrt{\rho} A-\Exp{A}_\rho^2 \Bigr)\Bigl( \tr \sqrt{\rho} B \sqrt{\rho} B -\Exp{B}_\rho^2\Bigr), \label{bddPL} \\
		&B_{\rm 1}(A,B,\rho) = \frac{\lm^2}{2\lM} \|[A,B]\|^2_\rho, \label{bdd1} \\
		&B_{\rm 2}(A,B,\rho) = \frac{\lm \lSm}{\lm + \lSm} \|[A,B]\|^2_\rho. \label{bdd2}
	\end{align}
\end{subequations}
We have $B_{\rm S}(A,B,\rho), B_{\rm LP}(A,B,\rho) \ge B_{\rm R}(A,B,\rho)$ and $B_{\rm 2}(A,B,\rho) \ge B_{\rm 1}(A,B,\rho)$. The last inequality is seen by considering the order $\lm \le \lSm \le \lM$.
However, the quality of these bounds depends on the choice of the physical quantities $A$ and $B$, as well as the state $\rho$. In order to conduct a fair comparison and observe the universal properties and trends, our strategy is to first compare the average bounds across all pairs of physical quantities by restricting to qubit systems (Sec.~\ref{AvOb}) and second to compare the average over all pairs of mutually unbiased observables for any finite quantum system (Sec.~\ref{AvMUOb}).

\subsection{Average bounds over all pair of observables in qubit systems}\label{AvOb}

In this section, we compare the bounds in \eqref{AllBdd} by averaging all pairs of observables for qubit systems.
Notice that, for qubit cases, the tighter bound \eqref{conj:BW} is already proven.
Specifically, by expanding observables $A$ and $B$ by Pauli matrices $\sigma_1,\sigma_2,\sigma_3$:
\begin{align}\label{BlAB}
	A = \sum_{i=1}^3 a_i \sigma_i, B = \sum_{i=1}^3 b_i \sigma_i,
\end{align}
with unit vectors\footnote{Here, normalizations are performed to eliminate non-essential uncertainties due to the magnitude of the operators. Additionally, note that the components of the identity operator are independent of the uncertainty, and therefore are disregarded.} ${\bm a}=(a_1,a_2,a_3), {\bm b}=(b_1,b_2,b_3) \in \R^3$, we average the bounds \eqref{AllBdd} uniformly over the set of pair $({\bm a},{\bm b})$ of all $3$-dimensional unit vectors, integrating with respect to the Haar measure on the unit sphere.

For a qubit state $\rho$, it is convenient to use the Bloch vector representation (see e.g. \cite{NC,KIMURA2003339}):
\begin{align}\label{Bl}
	\rho = \frac{1}{2}(\I + \sum_{k=1}^3 c_k \sigma_k),
\end{align}
where ${\bm c} =(c_1,c_2,c_3) $ lies in the Bloch ball, i.e., $|{\bm c}|\le 1$.
Using the algebra of Pauli matrices, $\sigma_i \sigma_j = \delta_{ij} \I + i \sum_{k=1}^3 \epsilon_{ijk} \sigma_k \ (i,j=1,2,3)$, the purity $P := \tr \rho^2$ ($\ge 1/2$) and $\lm,\lSm=\lM$ are easily calculated as $P = \frac{1}{2}(1+|{\bm c}|^2)$, $\lm = \frac{1-|{\bm c}|}{2},\lSm=\lM = \frac{1+|{\bm c}|}{2}$ so that
\begin{align}\label{lamP}
	\lm = \frac{1-\sqrt{2P-1}}{2}, \lSm=\lM = \frac{1+\sqrt{2P-1}}{2}.
\end{align}
Moreover, a direct computation gives
\begin{subequations}
	\begin{align}
		&B_{\rm R}(A,B,\rho) = |(\bm{a}\times\bm{b})\cdot \bm{c}|^2,\label{rob}\\
		&B_{\rm S}(A,B,\rho) = |(\bm{a}\times\bm{b})\cdot \bm{c}|^2 +|\bm a\cdot\bm b-(\bm a\cdot\bm c)(\bm b\cdot\bm c)|^2,\label{sch}\\
		& B_{\rm LP}(A,B,\rho) = |(\bm{a}\times\bm{b})\cdot \bm{c}|^2, \nonumber \\
		& {} + \Bigl(\sqrt{1-|\bm c|^2} + (1-|\bm c|^2-\sqrt{1-|\bm c|^2})\frac{(\bm a\cdot\bm c)^2}{|\bm c|^2}\Bigr), \nonumber \\
		&\qquad  \times \Bigl(\sqrt{1-|\bm c|^2} + (1-|\bm c|^2-\sqrt{1-|\bm c|^2})\frac{(\bm b\cdot\bm c)^2}{|\bm c|^2}\Bigr), \label{PLbdd}\\
		&B_{\rm 1}(A,B,\rho)=2\frac{P-\sqrt{2P-1}}{1+\sqrt{2P-1}}|\bm a\times\bm b|^2,\label{ours1}\\
		&B_{\rm 2}(A,B,\rho)=2(1-P)|\bm a\times\bm b|^2\label{ours2},
	\end{align}
\end{subequations}
where $\bm{a}\times\bm{b}$ is the cross product.
As a side remark, $\|[A,B]\|_\rho^2$, appearing in \eqref{bdd1} and \eqref{bdd2}, becomes independent of a state $\rho$ for qubit system and coincide with $\frac{1}{2}\|[A,B]\|^2 = 4|\bm a\times\bm b|^2$.
This property, however, does not generally extend to systems of dimension $d \geq 3$.

Finally, by using general formulas \eqref{eq:AVunif} given in Appendix \ref{app:Uav}, we obtain their averaged bounds as functions of purity $P$ of a state $\rho$:
\begin{subequations}\label{avBdd}
	\begin{align}
		\langle B_{\rm R}(A,B,\rho)\rangle_{\rm av} &= \frac{2}{9}(2P-1) := B_{\rm R}(P),\label{rob2}\\
		\langle B_{\rm S}(A,B,\rho)\rangle_{\rm av} &= B_{\rm R}(P) + \frac{2}{9}(2P^2-4P+3),\label{sch2}\\
		\langle B_{\rm LP}(A,B,\rho) \rangle_{\rm av} &= B_{\rm R}(P) + \frac{4}{9}\Bigl((1-P) +\sqrt{2(1-P)}\Bigr)^2, \label{PLbdd-2} \\
		\langle B_{\rm 1}(A,B,\rho)\rangle_{\rm av}&=\frac{4}{3}\frac{P-\sqrt{2P-1}}{1+\sqrt{2P-1}},\label{ours1-2}\\
		\langle B_{\rm 2}(A,B,\rho)\rangle_{\rm av}&=\frac{4}{3}(1-P)\label{ours2-2}.
	\end{align}
\end{subequations}

\begin{figure}[htbp]
	\centering
	\includegraphics[width=\columnwidth]{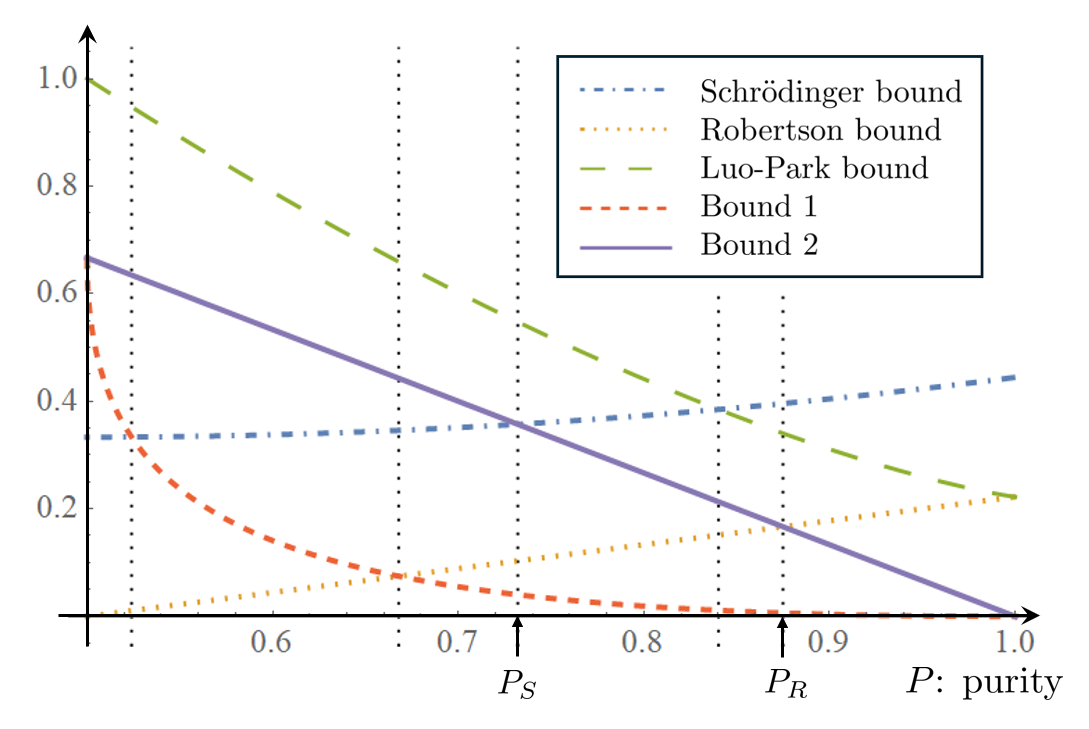}
	\caption{(Color Online) Average bounds in \eqref{avBdd} are plotted as a function of the purity $P$. The dashed (orange) line and the solid (purple) line are bounds \eqref{ours1-2} and \eqref{ours2-2}; the dotted (yellow) line, dash-dotted (blue) line and the long-dashed (green) line are the Robertson bound \eqref{rob2}, Schr\"odinger bound \eqref{sch2}, and Luo-Park bound \eqref{PLbdd-2}, respectively.}\label{fig:UR}
\end{figure}

Figure~\ref{fig:UR} shows the graphs of averaged bounds \eqref{rob2} - \eqref{ours2-2} as functions of the purity $P$.
One observes that, as the state becomes more mixed, both our bounds \eqref{ours1-2} and \eqref{ours2-2} surpass those of Robertson and Schr\"odinger.
Specifically, the bound \eqref{ours2-2} outperforms the Robertson bound if $P \le P_{\rm R}: = 7/8 =0.875$ and the Schr\"odinger bound if $P \le P_{\rm S} = \sqrt{3}-1 \simeq 0.732$.
This fact implies that our bounds are detecting a previously unidentified trade-off in quantum uncertainty arising from the non-commutativity of observables.

On the other hand, the LP bound exhibits a similar tendency to our bounds regarding the dependency on purity, and is outperforming them. However, this result is primarily an average behavior and, as seen below, our bound can provide a tighter bound for specific physical quantities.

\subsection{Average bounds over all pair of mutually unbiased observables}\label{AvMUOb}

Two non-degenerate observables $A$ and $B$ in $d$-level systems are said to be mutually unbiased (or complementary) if
$$
|\bracket{a_i|b_j}|^2 = \frac{1}{d} \quad \forall i,j = 1,\ldots, d
$$
where $\{\ket{a_i}\}_{i=1}^d$ and $\{\ket{b_j}\}_{j=1}^d$ are the set of normalized eigenvectors of $A$ and $B$.
Physically speaking, $A,B$ are mutually unbiased when one is most uncertain when the other is deterministic due to the Born's rule \cite{Schwinger,Ivonovic_1981,WOOTTERS1989363,Peres1995}. Therefore, the trade-off in uncertainty relations is expected to be most significant between physical quantities that are mutually unbiased.
This is indeed manifested in the entropic uncertainty relation \cite{Maassen}.
In this section, we compare our bounds and others for this important class of observables.

Before discussing the general cases, let us first consider some specific examples in qubit systems.
Let $A$ and $B$ be mutually unbiased and consider the case where $A$ and $\rho$ are commutative, i.e., the situation where a state has no coherence of $A$.
In qubit systems, these conditions correspond to the assumptions that the corresponding vectors $\bm a$ and $\bm b$ are perpendicular (for mutual unbiasedness), and ${\bm a}$ and ${\bm c}$ are parallel:
\begin{align}
	\bm a \perp \bm b, {\bm a} \parallel {\bm c}.
\end{align}
In this case, it is easy to see that both Robertson and Schr\"odinger bounds are zero (as ${\bm a}\times {\bm b} \perp {\bm c}$), and thus fail to capture the intrinsic complementarity within mutually unbiased observables.
On the other hand, both the LP bound \eqref{PLbdd} and our bounds \eqref{ours1} and \eqref{ours2} do, since we have
\begin{align*}
B_{\rm LP}(A,B,\rho) &= (2(1-P))^{3/2},\\
B_1(A,B,\rho) &= 2\frac{P-\sqrt{2P-1}}{1+\sqrt{2P-1}},\\
B_2(A,B,\rho) &= 2(1-P).
\end{align*}
Therefore, the bound \eqref{ours2} outperforms the LP bound for any $P$, while the bound \eqref{ours1} is inferior to the PL bound. In order to be slightly more general, consider mutually unbiased $A$ and $B$ (i.e., ${\bm a} \perp {\bm b}$), but let the state ${\bm c}$ lie in the plane composed of ${\bm a}$ and ${\bm b}$.
Choosing a Cartesian coordinate system where the $x$-axis and $y$-axis are oriented in the directions of ${\bm a}$ and ${\bm b}$, we can write ${\bm a}=(1,0,0), {\bm b}=(0,1,0), {\bm c} = \sqrt{2P-1}(\cos \theta, \sin \theta, 0)$ with a parameter $\theta \in [0,2\pi)$. Then, the LP bound \eqref{PLbdd} is calculated as
$$
q^2 (1 + (q-1) \cos^2 \theta)(1 + (q-1) \sin^2 \theta) = q^2(q+(q-1)^2\frac{\sin^2{2\theta}}{4})
$$
where $q = \sqrt{2(1-P)}$. Therefore, the LP bound takes the maximum when $\theta = \pi/4$:
$$
\frac{q^2(1+q)^2}{4}.
$$
and takes the minimum when $\theta = 0$, which is the above case.
Since our bound \eqref{ours2} remains to be the same $2(1-P) = q^2$, this is always greater than the PL bound for any $P$ and $\theta$.

Now, we consider general mutually unbiased observables $A$ and $B$ in $d$-level systems and treat the case where $A$ and $\rho$ are commutative: Let $\{\ket{j}\}_{j=1}^d$ be the common eigenbasis of $A$ and $\rho$ that simultaneously diagonalize them, so that their eigenvalue decompositions are given by $A = \sum_j a_j \ketbra{j}{j}$ and $\rho = \sum_j \lambda_j \ketbra{j}{j}$. Let $B = \sum_j b_j \ketbra{b_j}{b_j}$ be an eigenvalue decomposition of $B$, where we assume the mutually unbiased condition:
$$
|\bracket{j|b_k}|^2 = \frac{1}{d} \ \forall j,k = 1,\ldots,d.
$$
We can rewrite this condition using a phase information $\theta_{jk} \in \R$ by
\begin{align}\label{MUBphase}
	\bracket{j|b_k} =  \frac{1}{\sqrt{d}} e^{i \theta_{jk}}.
\end{align}
Notice that both Robertson and Schr\"odinger bounds vanish also in this general setting. This can be easily seen by using $[A,\rho] = 0$ and the cyclic property of trace as follows:
\begin{align*}
	B_{\rm R}(A,B,\rho) &= \frac{1}{4} |\tr [A,B]\rho |^2 = \frac{1}{4} |\tr B [\rho, A] |^2 = 0. \\
	B_{\rm S}(A,B,\rho) &= 0 + \Bigl|\frac{1}{2}\Exp{\{A,B\}}_\rho-\Exp{A}_\rho\Exp{B}_\rho\Bigl|^2 \nonumber \\
	& = \Bigl|\tr BA \rho -\tr A \rho \tr B \rho \Bigl|^2.
\end{align*}
However, the last term vanishes as
\begin{align*}
	&\tr BA \rho -\tr A \rho \tr B \rho \\\
	&= \sum_{j,k} a_j \lambda_j b_k |\bracket{j|b_k}|^2 - (\sum_i a_i \lambda_i) (\sum_{j,k} b_k \lambda_j |\bracket{j|b_k}|^2 ) \\
	& = \frac{1}{d} (\sum_j a_j \lambda_j) (\sum_k b_k)  - \frac{1}{d} (\sum_i a_i \lambda_i) (\sum_j \lambda_j )(\sum_{k} b_k ) \\
	& = 0.
\end{align*}
Therefore, we emphasize again that, in general, the Robertson and Schr\"odinger relations fail to capture the complementarity between mutually unbiased physical quantities.
On the other hand, both the LP bound \eqref{bddPL} and our bounds (\eqref{bdd1} and \eqref{bdd2}) are capable of detecting this complementarity, as the LP bound \eqref{bddPL} is given by the product of
\begin{align}\label{a0sqrMUB}
	\tr({A\sqrt{\rho}A\sqrt{\rho}}) - (\tr A \rho)^2 = \sum_j a^2_j \lambda_j - \sum_{j,k} a_j a_k \lambda_j \lambda_k,  \end{align}
and
\begin{align}\label{b0sqMUB}
	&\tr({B\sqrt{\rho}B\sqrt{\rho}}) - (\tr B \rho)^2 \nonumber \\
	&= \sum_{j,k} \sqrt{\lambda_j} \sqrt{\lambda_k}|\bracket{k|B |j}|^2 - (\sum_j \lambda_j \bracket{j|B |j})^2 \nonumber \\
	&= \frac{1}{d^2}\sum_{j,k} \sqrt{\lambda_j} \sqrt{\lambda_k} \sum_{l,m} b_l b_m e^{-i(\theta_{kl}-\theta_{jl})}e^{i(\theta_{km}-\theta_{jm})} \nonumber \\
	& \quad {} - \frac{1}{d^2} (\sum_{j,k} b_j b_k),
\end{align}
and our bounds \eqref{bdd1} (resp. \eqref{bdd2}) are given by the product of $\frac{\lm^2}{2\lM}$ (resp. $\frac{\lm \lSm}{\lm + \lSm} $) and the $\rho$-norm of the commutator:
\begin{align}\label{bdd2MUB}
	&\|[A,B]\|^2_\rho = \tr (B A^2 B \rho) + \tr (B^2 A^2 \rho) - 2 \tr (BABA \rho)  \nonumber \\
	&= \sum_{j,k} \lambda_k a_j (a_j - 2 a_k) |\bracket{j|B|k}|^2 + \sum_j \lambda_j a_j^2 \bracket{j|B^2 | j} \nonumber \\
	&= \frac{1}{d^2} \sum_{j,k} \lambda_k a_j (a_j - 2 a_k) \sum_{l,m} b_l b_m e^{-i(\theta_{kl}-\theta_{jl})}e^{i(\theta_{km}-\theta_{jm})} \nonumber \\
	& \quad {} + \frac{1}{d} \sum_j \lambda_j a_j^2 \sum_l b_l^2.
\end{align}

Now, we would like to compare the LP bound with our bounds, especially the tighter one \eqref{bdd2}.
In order to conduct a general comparison, we take the average over all eigenvalues over the sphere $S_{d-1} \subset \R^d$ of a mutually unbiased pair of $A$ and $B$.

For the LP bound, we can compute the averages of \eqref{a0sqrMUB} and \eqref{b0sqMUB} over ${\bm a} \in S_{d-1}$ and ${\bm b} \in S_{d-1}$ independently; Using the formula \eqref{eq:AVunif} in Appendix \ref{app:Uav}, it is easy to compute the average of \eqref{a0sqrMUB} and \eqref{b0sqMUB}, given respectively by
\begin{align}
	\sum_j \frac{1}{d} \lambda_j - \sum_{j,k} \frac{\delta_{jk}}{d} \lambda_j \lambda_k = \frac{1}{d} - \frac{1}{d} \sum_j \lambda_j^2 = \frac{1 - P}{d}
\end{align}
and
\begin{align}
	\frac{1}{d^2}\sum_{j,k} \sqrt{\lambda_j} \sqrt{\lambda_k} - \frac{1}{d^2} = \frac{(\sum_j \sqrt{\lambda_j})^2 - 1 }{d^2}.
\end{align}
Therefore, we obtain
\begin{align}\label{bdd:LPavMU}
	\langle B_{\rm LP}(A,B,\rho) \rangle_{{\rm av.}} = \frac{(1 - \sum_j \lambda^2_j)((\sum_j \sqrt{\lambda_j})^2 - 1)}{d^3}.
\end{align}

\begin{figure}[htbp]
	\centering
	\includegraphics[width=\columnwidth]{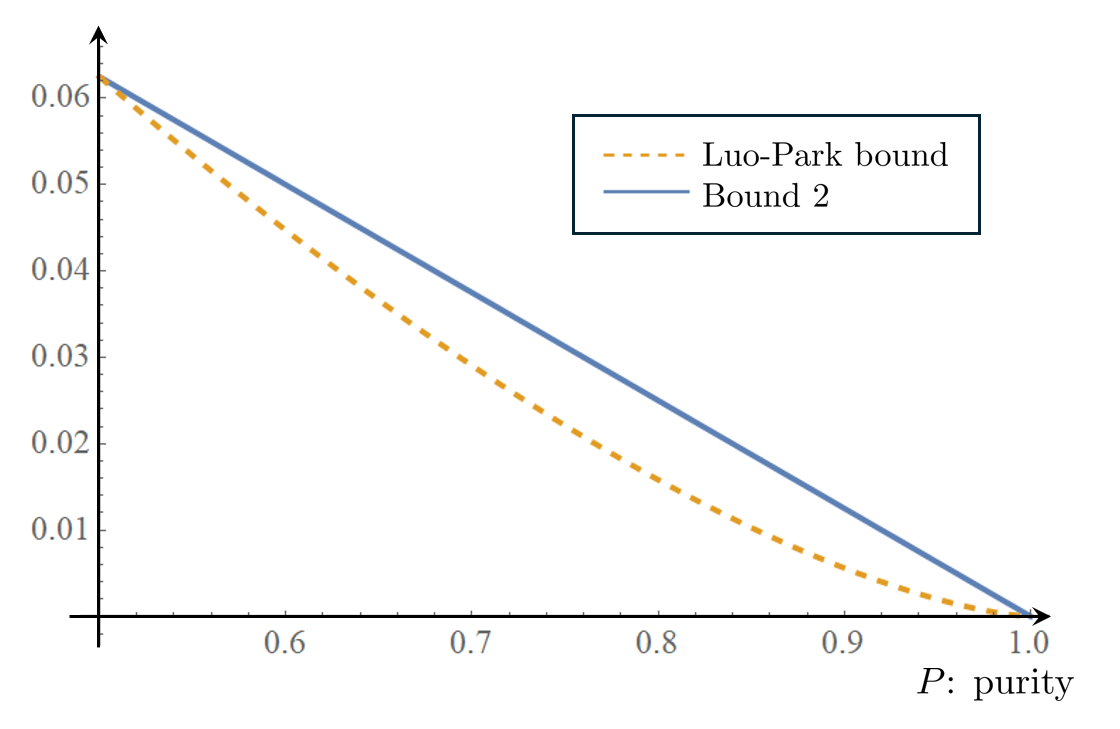}
	\caption{(Color Online) Average over mutually unbiased observables for Luo-Park bound \eqref{bdd:LPavMU} and our bound \eqref{bdd:OuravMU} are plotted in dashed (yellow) line and solid (blue) line, respectively. }\label{fig:2}
\end{figure}

To compute the average of our bound, using the formula \eqref{eq:AVunif} in Appendix \ref{app:Uav}, one can compute the average of \eqref{bdd2MUB} over ${\bm b}$ and then average further over ${\bm a}$ to obtain
\begin{align}
	\langle \|[A,B]\|^2_\rho \rangle_{{\rm av.}} = \frac{1}{d^2} - \frac{2}{d^3}  + \frac{1}{d^2} = \frac{2(d-1)}{d^3} = \left\{
	\begin{array}{cc}
		\frac{1}{4} & d= 2 \\
		\frac{4}{27} & d= 3 \\
		\frac{3}{32} & d= 4 \\
		:
	\end{array}
	\right.
	\nonumber \\
\end{align}
Therefore, we have
\begin{align}\label{bdd:OuravMU}
	\langle B_2(A,B,\rho) \rangle_{{\rm av.}} = \frac{\lm \lSm}{\lm + \lSm} \frac{2(d-1)}{d^3}.
\end{align}
In the case of general dimensions, whether \eqref{bdd:LPavMU} or \eqref{bdd:OuravMU} provide a better bound depends on the eigenvalues of the state. However, in the case of $d=2$, we find that our bound always outperform the LP bound (See Figure \eqref{fig:2}]).

\section{Conclusion}\label{S-IV}

In this paper, we have introduced two uncertainty relations based on generalizations of the B\"ottcher-Wenzel inequality utilizing state-dependent norms of commutators.
The first relation \eqref{newULoos2} is mathematically proven, while the second tighter relation \eqref{newU2} is strongly supported by extensive numerical evidence and proven for the case of qubit systems.
Through systematic comparisons, we have demonstrated that our uncertainty relations can surpass the conventional Robertson and Schr\"odinger bounds, especially as the quantum state becomes increasingly mixed. This reveals a previously unrecognized aspect of quantum uncertainty originating from the non-commutativity of observables.
On the other hand, the Luo-Park relation exhibits a similar dependence on state purity and outperforms our bounds on average over all observables in qubit systems. However, when considering the important case of mutually unbiased observables, our tighter relation \eqref{newU2} can surpass the Luo-Park bound, potentially providing a better description of the complementarity between such observables.
Overall, our results shed new light on the role of non-commutativity in quantum uncertainty relations and may find applications in fields where such trade-off relations play a fundamental role, such as quantum metrology and quantum computing.

\begin{acknowledgements}
	AM and GK would like to thank Jaeha Lee, Michele Dall'Arno and Izumi Tsutsui for the useful discussions and comments.
	GK and HO are supported by JSPS KAKENHI Grant Number 24K06873 and 23K03147, respectively.  
	DC was supported by the Polish National Science Center project No. 2018/30/A/ST2/00837.
\end{acknowledgements}

\appendix

\section{Proof of BW inequality for normal matrices}\label{app:BW}

Here, let us introduce a simple proof of the BW inequality \eqref{BW} given in \cite{Bottcher2005} for the cases where either $A$ or $B$ is normal: 

[Proof] Let $A$ be a normal matrix, i.e., $[A,A^\dagger] = 0$ and $A = \sum_{i=1}^d a_i \ketbra{a_i}{a_i}$ be an eigenvalue decomposition of $A$. Let also $b_{ij} = \bracket{a_i|B a_j}$.
A direct computation shows $\|[A,B]\|^2 = \tr [A,B]^\dagger [A,B] = \sum_{i,j}|a_i - a_j|^2 |b_{ij}|^2$. However, this is bounded from above by $\sum_{i \neq j}|a_i - a_j|^2 |b_{ij}|^2 \le \sum_{i \neq j}2(|a|^2_i + |a|^2_j) |b_{ij}|^2 \le 2 \sum_{i \neq j}\|A\|^2 |b_{ij}|^2 \le 2 \|A\|^2 \|B\|^2$, where we have used $\|A\|^2 = \sum_i |a_i|^2 $ and $\|B\|^2 =\sum_{i,j} |b_{ij}|^2$. \hfill $\blacksquare$


\section{Proof of BW inequality for qubit systems}\label{app:Pd2}

In this appendix, we present an elementary proof of \eqref{conj:BW} for $d=2$.
For our application, $\rho$ is a normalized density operator, and both $A$ and $B$ are assumed to be Hermitian. For the general proof, see \cite{MKHC}.

[Proof] Similar to \eqref{BlAB}, expand arbitrary Hermitian matrices $A,B$ with Pauli basis:
\begin{align}\label{BlAB2}
	A = a_0 \I + \sum_{i=1}^3 a_i \sigma_i, B = b_0 \I + \sum_{i=1}^3 b_i \sigma_i,
\end{align}
with real numbers $a_0, b_0$ and ${\bm a} = (a_1,a_2,a_3), {\bm b} = (a_1,a_2,a_3) \in {\mathbb R}^3$.
Using Bloch vector representation \eqref{Bl} for $\rho$, one has
\begin{align}
	\|[A,B]\|^2_\rho = 4 |{\bm a} \times {\bm b}|^2  \label{eq:ABcomRho}
\end{align}
\begin{align}
	\|A\|^2_\rho =a_0^2 + |{\bm a}|^2 + 2a_0 {\bm c} \cdot {\bm a}
\end{align}
\begin{align}
	\|B\|^2_\rho =b_0^2 + |{\bm b}|^2 + 2b_0 {\bm c} \cdot {\bm b}
\end{align}
and $\lambda_1 \lambda_2 = \frac{1-|{\bm c}|^2}{4}$.
Therefore, inequality \eqref{conj:BW} is equivalent to
$$
(a_0^2 + |{\bm a}|^2 + 2 a_0 {\bm c} \cdot {\bm a})(b_0^2 + |{\bm b}|^2 +2 b_0 {\bm c} \cdot {\bm b}) \ge (1-|{\bm c}|^2) |{\bm a} \times {\bm b}|^2
$$
for any $a_0,b_0 \in \R$ and ${\bm a},{\bm b},{\bm c} \in \R^3$ where $|{\bm c}|^2 \le 1$.
Since $(a_0^2 + |{\bm a}|^2 + 2 a_0 {\bm c} \cdot {\bm a}) = (a_0 +  {\bm c} \cdot {\bm a})^2  + |{\bm a}|^2 -  ({\bm c} \cdot {\bm a})^2 \ge |{\bm a}|^2 -  ({\bm c} \cdot {\bm a})^2$, and the same holds for $B$, it is enough to show
\begin{equation}\label{inab}
	(|{\bm a}|^2 - ({\bm c} \cdot {\bm a})^2)(|{\bm b}|^2 - ({\bm c} \cdot {\bm b})^2) \ge (1-|{\bm c}|^2) |{\bm a} \times {\bm b}|^2. 	
\end{equation}
By choosing $z$-axis in the direction of ${\bm c}$, one may write ${\bm c} = (0,0,r)$ for $0\le r \le 1$, without loss of generality, so that \eqref{inab} reads
\begin{align}\label{eq:A2}
	(|{\bm a}|^2 - r^2 a_3^2)(|{\bm b}|^2 - r^2 b_3^2) \ge (1-r^2) |{\bm a} \times {\bm b}|^2.
\end{align}
If we introduce compressed vectors ${\bm a}' = (a_1,a_2,\sqrt{(1-r^2)}a_3)$, ${\bm b}' = (b_1,b_2,\sqrt{(1-r^2)}b_3)$ with a factor $\sqrt{1-r^2}$ in $z$-direction, one can easily show
\begin{align}\label{eq:A}
	|{\bm a'} \times {\bm b'}|^2 \ge (1-r^2) |{\bm a} \times {\bm b}|^2.
\end{align}
(This can be readily shown by computing the vector components for the cross products.)
By noting $|{\bm a'}|^2 = (|{\bm a}|^2 - r^2 a_3^2), \ |{\bm b'}|^2 = (|{\bm b}|^2 - r^2 b_3^2)$ and $|{\bm a'}|^2 |{\bm b'}|^2 \ge |{\bm a'} \times {\bm b'}|^2$, inequality \eqref{eq:A} implies \eqref{eq:A2}.
\hfill $\blacksquare$

\section{Uniform average on the sphere}\label{app:Uav}

In this section, we give a useful formula for the average over the uniform measure $d\mu$ on the sphere $S_{d-1} \subset \R^d$. The average of the function $f({\bm x})$ of ${\bm x} \in \R^d$ is given by
\begin{align}
	\langle f({\bm x}) \rangle_{{\rm av.}} := \int_{S_{d-1}} d\mu f({\bm x}).
\end{align}
For any unit vector ${\bm x} = (x_1,\ldots,x_d) \in S_{d-1}$, we have
\begin{align}\label{eq:AVunif}
	\langle x_j x_k \rangle_{{\rm av.}} := \int_{S_{d-1}} d\mu x_j x_k = \frac{\delta_{jk}}{d}.
\end{align}
For completeness, we provide an elementary proof of this formula below.

[Proof] By the rotational symmetry, it is clear that $\int d\mu x^2_i $ does not depend on $i =1,\ldots,d$ and also for $i \neq j =1,\ldots,d$, $\int d\mu x_i x_j $ does not depend on the pair $i \neq j$: Let $c := \int d\mu x^2_i $ and $c' := \int d\mu x_i x_j \ (i \neq j)$. From the normalization condition,
$$
1 = \int_{S_{d-1}} d\mu \sum_{i} x^2_i = \sum_i \int_{S_{d-1}} d\mu x^2_i,
$$
and we get $\sum_i \int d\mu x^2_i = \frac{1}{d}$.
Next, observe that
$$
\int d\mu (\sum_i x_i)^2 = \sum_i \int d\mu x^2_i + \sum_{i \neq j} \int d\mu x_i x_j = 1 + (d^2-d) c'
$$
On the other hand, if we choose another coordinate system $x'_i$ satisfying $x'_1 = \frac{1}{\sqrt{d}}\sum_i x_i$, the left hand side is
$$
\int d\mu (\sqrt{d} x'_1)^2 = d \times \frac{1}{d} = 1.
$$
Therefore, we have $c' = 0$. \hfill $\blacksquare$

\bibliography{ref_UR}

\end{document}